\documentclass[english,12pt,aps,prd,a4paper,preprintnumbers,floatfix,nofootinbib,showpacs,notitlepage]{revtex4-1} 
\usepackage[mode=buildnew]{standalone}
\usepackage[usenames,dvipsnames]{color}  
\usepackage{graphicx}

\usepackage{setspace}
\usepackage{caption}
\usepackage{subcaption}
\usepackage{amsmath}
\usepackage{amssymb}
\usepackage[colorlinks=true,citecolor=darkred,urlcolor=darkred, pdfborder={0 0 0}]{hyperref}
\usepackage[normalem]{ulem}
\usepackage{xcolor}
\usepackage{placeins}
\usepackage{mathrsfs}
\usepackage{verbatim} 
\usepackage{cancel} 
\usepackage{float} 

\usepackage{scalerel}
\usepackage{tikz-feynman}
\tikzfeynmanset{compat=1.1.0}
\usepackage{feynmp}
\usepackage{tikzsymbols}
\usepackage{multirow}

\makeatletter
\def\p@subsection{}
\makeatother

\definecolor{darkred}{rgb}{0.6,0,0}

\definecolor{linkcolor}{rgb}{0,0,0.5}

\def\gsim{\raise0.3ex\hbox{$\;>$\kern-0.75em\raise-1.1ex\hbox{$\sim\;$}}}
\def\lsim{\raise0.3ex\hbox{$\;<$\kern-0.75em\raise-1.1ex\hbox{$\sim\;$}}}

\def\beqn#1{\begin{equation}\label{#1}}
\def\eeqn{\end{equation}}

\def\beqa#1{\begin{eqnarray}\label{#1}}
\def\eeqa{\end{eqnarray}}

\newcommand {\ignore}[1]{}

\def\Z4{$Z_4$}
\def\O5{$\mathcal{O}_5$ }

\def\321{$\mathrm{SU(3) \otimes SU(2) \otimes U(1)}$ }

\def\red{\color{red}{}}

\usepackage{orcidlink}
 
 \newcommand{\AddrSeoul}{Institute for Convergence of Basic Studies, Seoul National University of Science and Technology, Seoul 01811, Republic of Korea \\ \\  \href{mailto:kumarranjeet.drk@gmail.com}{\red{kumarranjeet.drk@gmail.com}}}

\begin{document}

\title{\color{BrickRed} Fate of $\boldsymbol{\theta_{12}}$ under $\boldsymbol{\mu-\tau}$ Reflection Symmetry in Light of the First JUNO Results}

\author{Ranjeet Kumar\orcidlink{0000-0002-7144-7606}
}
\affiliation{\AddrSeoul}

\begin{abstract}
  \vspace{1cm} 
\noindent
The recent JUNO measurements of $\theta_{12}$ and $\Delta m^2_{21}$ open a new avenue for probing flavor symmetric structures in the lepton sector. Motivated by this, we study a model in which $\mu-\tau$ reflection symmetry naturally emerges from an underlying $A_4$ flavor symmetry within a type-II seesaw framework. Beyond its standard predictions of $\theta_{23}=45^{\circ}$ and $\delta_{\rm CP}=\pm \pi/2$, the framework yields testable predictions for $\theta_{12}$ that can be probed by JUNO. Two viable scenarios arise, one predicting $\sin^2\theta_{12} \gsim 0.335$, which is strongly disfavored by the latest JUNO results. Correlations between $\theta_{12}$ and model parameters further enhance the model's predictivity. Future measurements at DUNE and T2HK will provide complementary tests of this scenario.
\end{abstract}

\maketitle


\section{Introduction} \label{sec:intro}

The precision era of neutrino oscillation physics has been marked by substantial progress in probing lepton mixing parameters~\cite{Kamiokande-II:1990wrs,Kamiokande-II:1992hns,Super-Kamiokande:1998kpq,Cleveland:1998nv,SNO:2002tuh,DoubleChooz:2011ymz,DayaBay:2012fng,RENO:2012mkc}. 
Recently, the next-generation Jiangmen Underground Neutrino Observatory (JUNO) has released its first measurements based on reactor antineutrino data~\cite{JUNO:2021vlw,JUNO:2022mxj,JUNO:2024jaw}. With only 59.1 days of exposure, JUNO has already achieved unprecedented sensitivity to the solar oscillation parameters, $\sin^2\theta_{12}$ and $\Delta m^2_{21}$ \cite{JUNO:2025gmd}.
The reported best-fit values with $1\sigma$ uncertainties for the normal mass ordering scenario are 
\begin{align}
    \sin^2{\theta_{12}}&=0.3092 \pm 0.0087, \nonumber \\ \Delta m^2_{21}&=(7.50 \pm 0.12)\times 10^{-5} \ \text{eV}^2 \ .
\end{align}
Even with its initial dataset, JUNO provides tighter constraints than current global fit analyses~\cite{ParticleDataGroup:2024cfk}, further supported by independent reactor measurements from the SNO+ collaboration~\cite{SNO:2025chx}.
In turn, the solar sector is now strongly constrained, enabling more incisive tests of lepton mixing frameworks and placing tighter limits on viable flavor models~\cite{King:2017guk,Feruglio:2019ybq,Xing:2020ijf,Almumin:2022rml,Chauhan:2023faf,Ding:2023htn,Ding:2024ozt}.
Nonetheless, several fundamental questions remain to be addressed.

At present, the key unknowns in the neutrino oscillation sector are the neutrino mass ordering i.e., the sign of $\Delta m^2_{31}$ ($\Delta m^2_{31}>0$ corresponds to normal ordering and $\Delta m^2_{31}<0$ to inverted ordering), the octant of the atmospheric angle $\theta_{23}$ (lower octant if $\theta_{23} < 45^{\circ}$ and upper octant if $\theta_{23} > 45^{\circ}$), and the Dirac CP-violating phase $\delta_{\rm CP}$. 
The first JUNO results~\cite{JUNO:2025gmd} already establish the foundation for its long term program to determine the neutrino mass ordering, while the remaining unknowns ($\theta_{23}$ and $\delta_{\rm CP}$) will be probed with high sensitivity by upcoming long-baseline experiments such as DUNE~\cite{DUNE:2016hlj,DUNE:2016rla,DUNE:2021cuw} and T2HK~\cite{Hyper-KamiokandeProto-:2015xww,Hyper-Kamiokande:2016srs}.
The release of the first JUNO data have motivated a wide range of theoretical studies~\cite{Chao:2025sao,Li:2025hye,Huang:2025znh,Ge:2025cky,Xing:2025xte,Chen:2025afg,Petcov:2025aci,Goswami:2025wla,Ding:2025dqd,Borah:2025vtn,Nanda:2025fvw,Shang:2026qkh}, including frameworks that predict nontrivial correlations among lepton mixing parameters \cite{Zhang:2025jnn,He:2025idv,Jiang:2025hvq,Ding:2025dzc,Dutta:2026dzh}. Consequently, such scenarios are now being subject to increasingly stringent constraints from precision measurements in the solar sector.

In this context, we explore a flavor symmetric framework in which the precise JUNO measurements \cite{JUNO:2025gmd} play a decisive role, with further scrutiny expected from future long-baseline experiments~\cite{DUNE:2016hlj,DUNE:2016rla,DUNE:2021cuw,Hyper-KamiokandeProto-:2015xww,Hyper-Kamiokande:2016srs}. 
We focus on a realization of $\mu-\tau$ reflection symmetry, which conventionally predicts a maximal atmospheric mixing angle, $\theta_{23}=45^{\circ}$ and a maximal Dirac CP-violating phase, $\delta_{\rm CP}=\pm \pi/2$~\cite{Harrison:2002et}.
In its canonical formulation~\cite{Chen:2015siy,Nath:2018hjx,Chakraborty:2018dew,Nath:2018xih,Nath:2018xkz,Nath:2018zoi,Liao:2019qbb,Huang:2020kgt,Yang:2020qsa,Zhao:2023vkb,Shao:2024fem,Yang:2025yst}, the symmetry strictly fixes $\theta_{23}$ and $\delta_{\rm CP}$, while leaving the reactor mixing angle $\theta_{13}$ and solar mixing angle $\theta_{12}$ unconstrained, allowing them to take values consistent with experimental data.
In the realization considered here, the framework further imposes a nontrivial constraint on $\theta_{12}$, thereby enabling a direct confrontation with JUNO's precision results~\cite{JUNO:2025gmd}.

In this work, we present a framework in which $\mu-\tau$ reflection symmetry emerges naturally from an underlying $A_4$ flavor symmetry\footnote{A number of selected studies based on $A_4$ in the literature can be found in Refs.~\cite{Babu:2002dz,Chen:2005jm,Altarelli:2005yx,Borah:2017dmk,CentellesChulia:2017koy,Borah:2018nvu,Ding:2020vud,RickyDevi:2023fqd,CentellesChulia:2023osj,Kumar:2023moh,Singh:2024imk,Kumar:2024zfb,Nomura:2024atp,Palavric:2024gvu,Moreno-Sanchez:2025bzz,Kumar:2025cte,Kumar:2025zvv,CentellesChulia:2025bcg}.}. Thanks to the $A_4$ structure~\cite{Ishimori:2010au}, the framework establishes a predictive relation among the parameters, resulting in a constrained value of the solar mixing angle $\theta_{12}$.
The model is realized within a type-II seesaw mechanism~\cite{Schechter:1980gr,Magg:1980ut,Cheng:1980qt,Mohapatra:1980yp} incorporating two types of $SU(2)_L$ triplet scalars, $\Delta$ and $\Delta'$. The specific choice of the vacuum expectation value (vev) of these scalars gives rise to two different cases of $\mu-\tau$ reflection symmetry, each leading to distinctive constraints on the lepton mixing parameters.
In light of the recent JUNO results~\cite{JUNO:2025gmd}, we focus here on the exact $\mu-\tau$ reflection symmetry scenario, which, in addition to the canonical predictions for $\theta_{23}$ and $\delta_{\rm CP}$, also constrains the solar mixing angle $\theta_{12}$. This notable prediction is a direct consequence of the underlying correlations among the model parameters governing the neutrino mass matrix.
Possible deviations from this exact scenario would result in a non-maximal $\theta_{23}$ and determine its octant. Such scenarios could be probed by upcoming experiments such as DUNE~\cite{DUNE:2016hlj,DUNE:2016rla,DUNE:2021cuw} and T2HK~\cite{Hyper-KamiokandeProto-:2015xww,Hyper-Kamiokande:2016srs}, but they lie beyond the scope of the present study.

The structure of this paper is as follows. In Sec.~\ref{sec:model},  we outline the theoretical framework of the model based on $A_4$ flavor symmetry and discuss the emergence of a $\mu-\tau$  reflection symmetric neutrino mass matrix. In Sec. \ref{sec:numerical}, we present the numerical results of the model, including possible correlations involving the mixing angle $\theta_{12}$ and examine their consistency with the recent JUNO results. Finally, we provide concluding remarks in Sec.~\ref{sec:conc}. In App. \ref{app:stu_para}, we provide a brief discussion of the corrections to the oblique parameters. The $A_4$ algebra and its tensor product rules are given in App. \ref{app:A4}. In App. \ref{app:lepmix}, we discuss the determination of lepton mixing parameters.

\section{Model Framework} 
\label{sec:model}

We consider a type-II seesaw model embedded within an $A_4$ flavor symmetry, in which $\mu-\tau$ reflection symmetry arises naturally. Neutrino masses are generated via the type-II seesaw mechanism, and the $A_4$ symmetry governs the leptonic mixing pattern.
We introduce two types of $SU(2)_L$ triplet scalars, $\Delta_i$ and $\Delta'$, transforming as $A_4$ singlets ($1,1'',1'$) and a triplet ($3$), respectively\footnote{The corrections to the oblique parameters~\cite{Peskin:1991sw}, induced by the presence of multiple ${SU}(2)_L$ triplet scalars, are briefly discussed in App.~\ref{app:stu_para}.}. The leptons $L$ and $l_{R_i}$ ($i=1,2,3$) are assigned as $3$ and $(1,1'',1')$ under $A_4$, respectively. The Higgs-like doublets $\phi_{\alpha}$ ($\alpha=e,\mu,\tau$), responsible for generating charged lepton masses, are assigned as triplets under $A_4$. In addition, a discrete $Z_3$ symmetry is imposed to ensure that the charged lepton mass matrix remains diagonal, so that the leptonic mixing pattern is solely governed by the neutrino sector. Under the $Z_3$ symmetry, non-trivial charges are assigned exclusively to $l_{R_i}$ and $\phi_{\alpha}$, whereas all other particles transform trivially. The particle content and their transformation properties under the various symmetries are summarized in Table \ref{tab:particles}. 
\begin{table}[!h]
    \centering
    \begin{tabular}{|c||c|c|c| c|}
        \hline 
        \ Fields \ &  \ $SU(2)_L$ \ & \ $U(1)_Y$ \ & \ $ A_4$ ~ & ~ $\mathbf{Z_3}$  \ \\ \hline
         $L$ &   $2$ & $-1/2$ & $3$ & $\mathbf{1}$  \\
         ${l_{R_i}}$ &  $1$ & $-1$ & $(1, 1'', 1')$ & ~ $\boldsymbol{(1, \omega^2, \omega)}$ ~  \\ 
         $\phi_{\alpha}$ &  $2$ & $1/2$ & $(3,3,3)$ & $\boldsymbol{(1, \omega, \omega^2)}$   \\
         $\Delta_i$ &  $3$ & $1$ & $(1, 1'', 1')$ & $\mathbf{1}$  \\
         $\Delta'$  & $3$ & $1$ & $3$  & $\mathbf{1}$
         \\ 
         \hline 
    \end{tabular}
    \caption{Particle content and their transformation properties under different symmetries, where $i=1,2,3$, and $\alpha=e,\mu,\tau$. To avoid confusion, the $\boldsymbol{\omega}$ charge associated with the $Z_3$ symmetry is indicated in \textbf{boldface}, distinguishing it from the $\omega$ that appears in the $A_4$ tensor product rules (see App. \ref{app:A4}) and vev alignment.}
    \label{tab:particles}
\end{table}

The transformations of $\Delta_i$ and $\Delta'$ are chosen such that their vev alignments $\langle \Delta_i \rangle = u_i$ and $\langle \Delta' \rangle =\frac{u'}{\sqrt{3}} (\pm 1, \omega, \omega^2)$ naturally realize the $\mu-\tau$ reflection symmetry. Depending on the sign choice in the vev alignment of $\Delta'$, two distinct scenarios arise.
Consequently, the model predicts the salient features of $\mu-\tau$ reflection symmetry, namely $\theta_{23} = 45^{\circ}$ and $\delta_{\rm CP}=\pm \pi/2$. Beyond the standard $\mu-\tau$ reflection predictions, the structure imposed by the underlying $A_4$ symmetry tightly correlates the model parameters, resulting in robust and testable predictions for the solar mixing angle $\theta_{12}$.
Imposing neutrino oscillation constraints yields a lower bound on $\sin^2\theta_{12}$ in one scenario, thereby enabling the latest JUNO results to probe this realization of the model. Having discussed the model framework, we next examine the generation of neutrino masses and the resulting mixing pattern.


\subsection{Lagrangian and Neutrino Mass Matrix} \label{sec:lag}

Following the charge assignments of the particles under the various symmetries given in Table \ref{tab:particles}, the invariant Yukawa Lagrangian that governs the leptonic sector can be formulated as follows
\begin{align} \label{eq:lag}
    \mathcal{L}= \mathcal{L}_{\ell} \ +  \ \mathcal{L}_{\nu}, \end{align}
where $\mathcal{L}_{\ell}$ is responsible for generating the charged lepton masses, and the neutrino masses can be extracted from $\mathcal{L}_{\nu}$, given by
\begin{align} \label{eq:lag2}
    -\mathcal{L}_{\ell}= & ~ y_e \left( \bar{L} \otimes \phi_e \right)_\mathbf{1} \otimes \left(l_{R_1}\right)_{\mathbf{1}} +y_{\mu} \left( \bar{L} \otimes \phi_{\mu} \right)_{\mathbf{1'}} \otimes \left(l_{R_2}\right)_{\mathbf{1''}} + y_{\tau} \left( \bar{L} \otimes \phi_{\tau} \right)_{\mathbf{1''}} \otimes \left(l_{R_3}\right)_{\mathbf{1'}}  + \text{H.c.} \ , \nonumber \\
    -\mathcal{L}_{\nu}= ~ & \alpha_1 \left( \bar{L}^c \otimes L\right)_{\mathbf{1}} \otimes \left(i \tau_2 \Delta_1 \right)_{\mathbf{1}}  
    +\alpha_2 \left( \bar{L}^c \otimes L\right)_{\mathbf{1'}} \otimes \left(i \tau_2 \Delta_2 \right)_{\mathbf{1''}}  
    +\alpha_3 \left( \bar{L}^c \otimes L\right)_{\mathbf{1''}} \otimes \left(i \tau_2 \Delta_3 \right)_{\mathbf{1'}}  
    \nonumber \\
   +& ~ \beta \left( \bar{L}^c \otimes L\right)_{\mathbf{3S}} \otimes \left(i \tau_2 \Delta' \right)_{\mathbf{3}} + \text{H.c.} \ ,
\end{align}
 The lower indices in parentheses $[...]_p$; $p = 1, 1',1'', 3_S, 3$ denote the $A_4$ transformation of enclosed fields.
 The vev alignments for $\phi_{\alpha}$ are chosen as 
 \begin{align}
     \phi_e = v_1 (1,0,0)^T, \quad  \phi_{\mu} = v_2 (0,1,0)^T,  \quad  \phi_{\tau} = v_3 (0,0,1)^T \ .
 \end{align}
This leads to a diagonal charged lepton mass matrix, implying that the leptonic mixing arises solely from the neutrino sector.
We now rewrite $\mathcal{L}_{\nu}$ in its expanded form using the $A_4$ tensor product rules,
\begin{align} \label{eq:lag2}
    -\mathcal{L}_{\nu}= &\alpha_1 \left(\bar{L}^c_1 L_1 + \bar{L}^c_2 L_2 + \bar{L}^c_3 L_3  \right) i \tau_2 \Delta_1 + \alpha_2 \left( \bar{L}^c_1 L_1 + \omega \bar{L}^c_2 L_2 + \omega^2 \bar{L}^c_3 L_3  \right) i \tau_2 \Delta_2  \nonumber \\ +&\alpha_3 \left( \bar{L}^c_1 L_1 + \omega^2 \bar{L}^c_2 L_2 + \omega \bar{L}^c_3 L_3  \right) i \tau_2 \Delta_3 \nonumber \\
    +&\beta \left[ \left( \bar{L}^c_2 L_3 +  \bar{L}^c_3 L_2 \right) i \tau_2 \Delta'_a + \left( \bar{L}^c_3 L_1 +  \bar{L}^c_1 L_3 \right) i \tau_2 \Delta'_b + \left( \bar{L}^c_1 L_2 +  \bar{L}^c_2 L_1 \right) i \tau_2 \Delta'_c  \right]+ \text{H.c.} .
\end{align}
Here, the scalar $\Delta'$, transforming as a triplet under $A_4$, is defined as $\Delta' \equiv (\Delta'_a, \Delta'_b, \Delta'_c)^T$. Once the scalar fields $\Delta_i$ and $\Delta'$ acquire vevs, neutrino masses are generated. We adopt a specific vev alignments consistent with the minimization conditions of the scalar potential~\cite{Degee:2012sk,Carrolo:2022oyg,deMedeirosVarzielas:2025byb}, given by\footnote{Another possible choice of vev alignment is $\langle \Delta' \rangle = \frac{u'}{\sqrt{3}} (\pm 1, \eta, \eta^{-1})^T$, where $\eta^2 = \omega$, which leads to identical predictions and is phenomenologically equivalent.}.
\begin{align} \label{eq:vev}
    \langle \Delta_i \rangle = u_i,  \quad \langle \Delta' \rangle = \frac{u'}{\sqrt{3}} (\pm1, \omega, \omega^2)^T,  
\end{align}
where $\omega^3 =1$. We choose the couplings $\alpha_i, \beta$ and the vevs $u_i, u'$ to be real, so that the CP-violation in the leptonic sector originates solely from the $\omega$ appearing in the $A_4$ tensor product rules. Thus, using Eqs.~\eqref{eq:lag2} and \eqref{eq:vev}, the neutrino mass matrix can be extracted, which exhibits the $\mu-\tau$ reflection symmetric structure, given by
\begin{align} \label{eq:numatrix}
\mathcal{M}_{\nu}= \begin{pmatrix}
    A & C & C^{\ast} \\
    C & B& D \\
    C^{\ast} & D & B^{\ast} 
\end{pmatrix} \ ,
\end{align}
\begin{align} \label{eq:par}
    \text{where,} \quad A= \sum^3_{i=1} \alpha_i u_i, \quad B= \sum^3_{i=1} \omega^{(2+i)} \alpha_i u_i, \quad D = \pm  \frac{\beta u'}{\sqrt{3}},  \quad C = \omega^2 \frac{\beta u'}{\sqrt{3}} \equiv (\pm) D \omega^2 . 
\end{align}
The parameters $A$ and $D$ are real, whereas $B$ and $C$ are complex. 
As indicated in Eq.~\eqref{eq:par}, the parameters $C$ and $D$ are not independent, but are related through $C=(\pm)\omega^2 D$. Moreover, the parameters $A$ and $B$ are also not entirely arbitrary; rather, they arise from similar combinations of the underlying couplings and vevs, reflecting a common structural origin. These relations are a direct consequence of the underlying $A_4$ symmetry together with the specific vev alignments of $\Delta_i, \Delta'$. As a result, the neutrino mass matrix in Eq.~\eqref{eq:numatrix} is left with only four effective free parameters. This reduces the parameter freedom of the mass matrix compared to a completely general realization of $\mu-\tau$ reflection symmetry \cite{Harrison:2002et}. 

This reduced number of independent parameters has important implications for the flavor structure of leptonic mixing, specifically for the mixing angle $\theta_{12}$. In the conventional $\mu-\tau$ reflection symmetry, $\theta_{12}$ remains unconstrained, as the neutrino mass matrix contains sufficient freedom among its elements. However, in our framework, the additional structure imposed by the $A_4$ symmetry and vev alignments induces nontrivial correlations among the matrix elements. Since the solar mixing angle is particularly sensitive to the relative structure of the (1-2) sector of the mass matrix, these relations translate into correlations involving $\theta_{12}$. As we will discuss in the next section, these correlations manifest differently in two distinct scenarios once experimental constraints are imposed.
While both scenarios exhibit correlated behavior of $\theta_{12}$ with the model parameters, case-II leads to a more restricted parameter space, resulting in a lower bound on 
$\sin^2{\theta_{12}}$, whereas case-I allows a broader range consistent with current data. Therefore, the predictivity of $\theta_{12}$ in this framework arises from the interplay between the $A_4$ imposed structure, the specific vacuum alignments, and the experimental constraints.
%


\section{Numerical Analysis and Model Predictions} 
\label{sec:numerical}

We now present a detailed numerical analysis of the model based on the neutrino mass matrix obtained in Eq.~\eqref{eq:numatrix}, which exhibits an exact $\mu-\tau$ reflection symmetry. Since the charged lepton mass matrix is diagonal, the leptonic mixing arises solely from the neutrino sector. As a consequence of the $\mu-\tau$ reflection symmetry, the atmospheric mixing angle and the Dirac CP-violating phase are fixed to $\theta_{23}=45^\circ$ and $\delta_{\rm CP}=\pm \pi/2$, respectively, independent of the specific values of the model parameters.
%
In light of the recent JUNO results~\cite{JUNO:2025gmd}, we focus on observables that play a decisive role in testing the framework. In particular, the correlated behavior of the solar parameters $\sin^2\theta_{12}$ and $\Delta m^2_{21}$ imposes strong constraints on the model.

We discuss two viable scenarios corresponding to the two possible vev alignments of the scalar $\Delta'$. The associated vev alignments are: $\langle \Delta' \rangle \propto (1,\omega,\omega^2)^T$ and $\langle \Delta' \rangle \propto (-1,\omega,\omega^2)^T$, which lead to $D=+\beta \frac{ u'}{\sqrt{3}}$ (case-I) and $D=-\beta \frac{ u'}{\sqrt{3}}$ (case-II), respectively, in Eq.~\eqref{eq:numatrix}.
The model involves four free parameters: two real parameters $A$ and $D$, and one complex parameter $B$, parametrized as $B = r e^{i \theta}$. We carry out a parameter scan over the following ranges:
\begin{align} \label{eq:range}
    A = \left[10^{-4}, 10^{-1} \right] \text{eV}, \quad |D| = \left[10^{-4}, 10^{-1} \right]\text{eV}, \quad r = \left[10^{-4}, 10^{-1} \right]\text{eV}, \quad  \theta = \left[0, 2 \pi \right], 
\end{align}
For the numerical analysis, we impose the $3\sigma$ constraints on neutrino oscillation parameters from the AHEP global fit \cite{deSalas:2020pgw,Rodejohann:2011vc}, given by
\begin{align} 
    &\Delta m^2_{21} = [6.94,8.14] \times 10^{-5} \ \text{eV}^2, \quad  \Delta m^2_{31} = [2.47,2.63] \times 10^{-3} \ \text{eV}^2, \nonumber \\
    & \sin^2{\theta_{12}}=[0.271,0.369], \quad \sin^2{\theta_{13}}=[0.02000,0.02405].
\end{align}
As noted above, the mixing angle $\theta_{23}$ is predicted to be $45^\circ$, which lies within the $3\sigma$ allowed range of AHEP global fit data.

We next present our model predictions for both scenarios mentioned above. The diagonalization procedure for the mass matrix in Eq.~\eqref{eq:numatrix}, together with the determination of the associated mixing angles, is outlined in App.~\ref{app:lepmix}.
We find that inverted ordering is disfavored in case-II, while case-I remains consistent with the current constraints. In the following, we focus on the normal ordering case, as it is the scenario probed by the recent JUNO measurements. The solar mixing angle $\theta_{12}$ exhibits a robust correlation with the model parameters in both scenarios.
For case-I, the predicted values of $\theta_{12}$ remain consistent with current data. In contrast, case-II predicts a lower bound on the solar mixing angle, $\sin^2\theta_{12} \gtrsim 0.335$, which is nearly excluded at the $3\sigma$ level considering the JUNO results~\cite{JUNO:2025gmd}, although it remains compatible with the AHEP global fit data~\cite{deSalas:2020pgw}.

\subsection{Constraints on Model Parameters from the Solar Mixing Angle $\boldsymbol{\theta_{12}}$}
\begin{figure}[h!]
    \centering
    \includegraphics[width=0.48\linewidth]{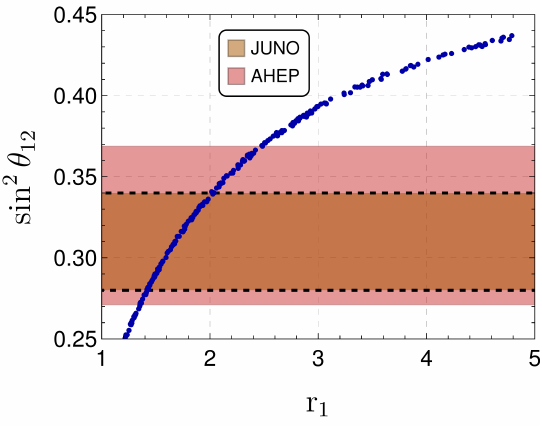}
     \includegraphics[width=0.48
     \linewidth]{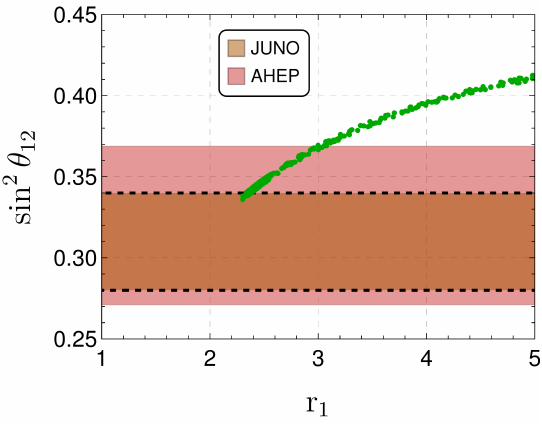}
    \caption{Correlation between $\sin^2\theta_{12}$ and the model parameter ratio $r_1$. The predictions for case-I and case-II are shown in blue and green, respectively, in the left and right panels. The brown and light red bands represent the allowed $3\sigma$ ranges from JUNO and AHEP, respectively.}
    \label{fig:r1}
\end{figure}

We begin by examining the correlation between the solar mixing angle $\theta_{12}$ and the model parameters. We find that $\sin^2\theta_{12}$ exhibits a strong dependence on the ratios of model parameters, defined as $r_1 \equiv |D/A|$, $r_2 \equiv |D/B|$, and $r_3 \equiv |A/B| \equiv r_2/r_1$. The allowed parameter spaces of the model are shown as blue and green scatter points corresponding to case-I and case-II, respectively. In Fig.~\ref{fig:r1}, we show the correlation between $\sin^2{\theta_{12}}$ and the ratio $r_1$.
The left and right panels correspond to case-I and case-II, respectively. The allowed $3 \sigma$ ranges of $\sin^2{\theta_{12}}$ from JUNO and AHEP global fit are indicated by the brown and light red bands, respectively.

\begin{figure}[!h]
    \centering
    \includegraphics[width=0.48\linewidth]{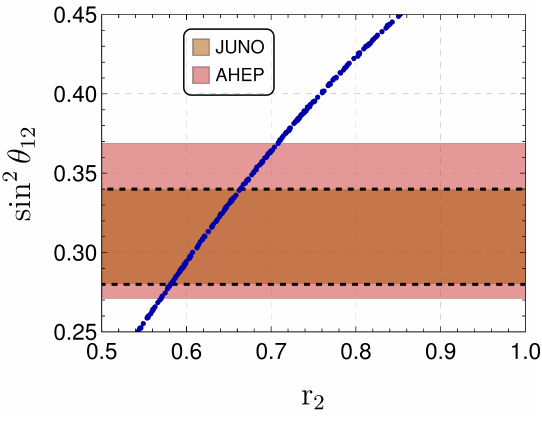}
    \includegraphics[width=0.48\linewidth]{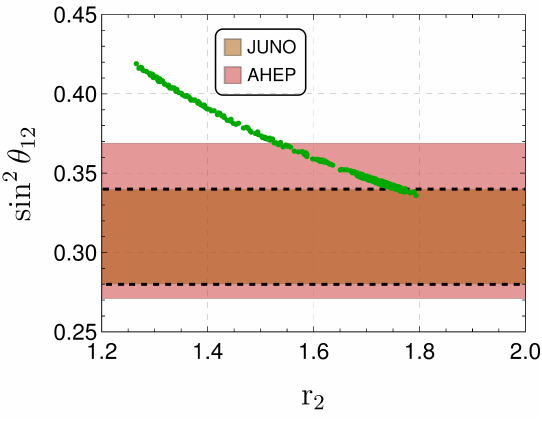}
    \caption{Correlation between $\sin^2\theta_{12}$ and the model parameter ratio $r_2$. The color code remains the same as in Fig.~\ref{fig:r1}.}
    \label{fig:r2}
\end{figure}

\begin{figure}[!t]
    \centering
    \includegraphics[width=0.48\linewidth]{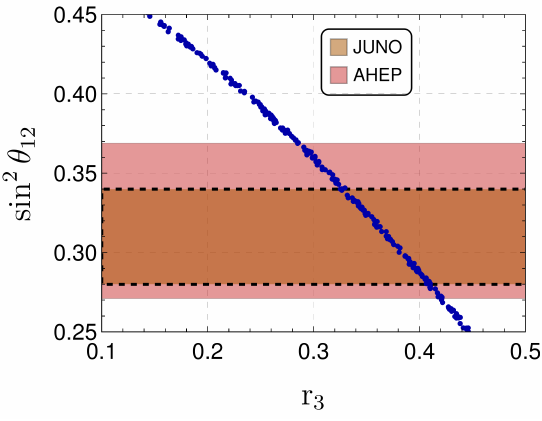}
    \includegraphics[width=0.48\linewidth]{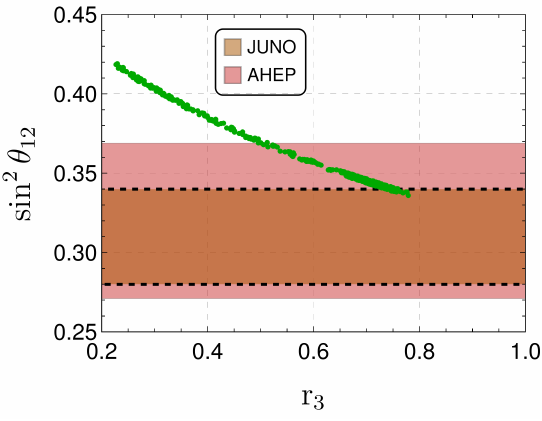}
    \caption{Correlation between $\sin^2\theta_{12}$ and the model parameter ratio $r_3$. The color code remains the same as in Fig.~\ref{fig:r1}.}
    \label{fig:r3}
\end{figure}

Similarly, we present the correlations for the ratios $r_2$ and $r_3$ in Figs.~\ref{fig:r2} and \ref{fig:r3}, respectively. The color code remains the same as in Fig.~\ref{fig:r1}.
The correlations observed in Figs.~\ref{fig:r1}, \ref{fig:r2}, and \ref{fig:r3} highlight the interplay between $\sin^2\theta_{12}$ and the parameter ratios. In view of the recent JUNO measurement, these relations translate into stringent tests of the model.
%
 Once the mass-squared differences ($\Delta m^2_{21}$ and $\Delta m^2_{31}$) constraints are imposed, the free parameters of the model become correlated, leading to fixed ratios. The additional requirement of compatibility with $\sin^2\theta_{12}$ further confines these ratios to a narrow region.
 
In case-I, imposing the $3\sigma$ AHEP constraints confines the parameter ratios to the ranges: $r_1 \sim [1.3,2.4]$, $r_2 \sim [0.57,0.70]$, and $r_3 \sim[0.28, 0.42]$. While, for case-II the corresponding allowed ranges are: $r_1 \sim [2.4,3.0]$, $r_2 \sim [1.5,1.8]$, and $r_3 \sim[0.50, 0.78]$.
In addition, case-II predicts a lower bound on the solar mixing angle, $\sin^2{\theta_{12}}\gsim 0.335$. As a result, this scenario survives only within a very narrow region of the AHEP global fit parameter space and is almost excluded by the JUNO measurements. Notably, the origin of this constraint lies in the intrinsic correlation among the solar parameters, which we discuss next.


\subsection{Solar Parameter Correlations in the Light of JUNO Results}

\begin{figure}[!h]
    \centering
    \includegraphics[width=0.48\linewidth]{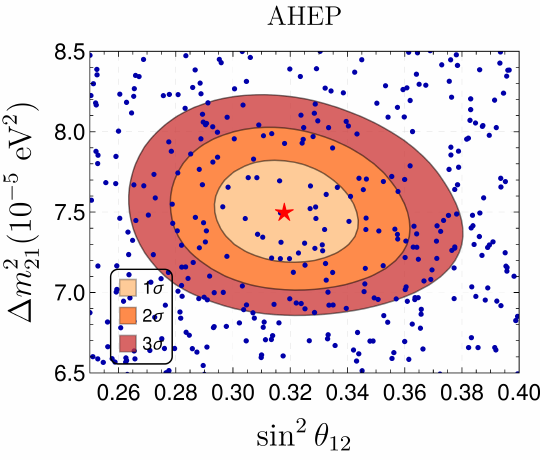}
     \includegraphics[width=0.48\linewidth]{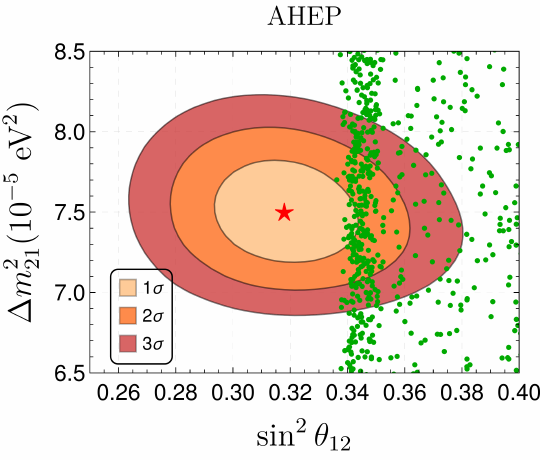}
    \caption{Correlation between the solar parameters $\sin^2\theta_{12}$ and $\Delta m^2_{21}$ predicted by the model, compared with the AHEP global fit data~\cite{deSalas:2020pgw}. Case-I and case-II are shown in blue and green in the left and right panels, respectively.}
    \label{fig:th12m21_AHEP}
\end{figure}

We now turn to the correlation between the solar parameters $\sin^2\theta_{12}$ and $\Delta m^2_{21}$. The resulting predictions can be directly compared with the AHEP global fit data~\cite{deSalas:2020pgw} and the JUNO measurements~\cite{JUNO:2025gmd}.
Figure~\ref{fig:th12m21_AHEP} illustrates the model correlations in the ($\sin^2{\theta_{12}}-\Delta m^2_{21}$) plane. The left panel represents the model predictions for case-I in blue, whereas the right panel shows the corresponding predictions for case-II in green.
For the comparison, we consider the $1,2,3\sigma$ contours from the AHEP global fit in the ($\sin^2{\theta_{12}}-\Delta m^2_{21}$) plane, with the corresponding best-fit point indicated by a red star. After imposing the constraints from the reactor mixing angle $\theta_{13}$ and the atmospheric mass-squared difference $\Delta m^2_{31}$, we find that the allowed parameter space in case-I, shown by blue points, spans the entire allowed region. In contrast, the parameter space in case-II, shown by green points, exhibits a lower bound on the solar mixing angle, $\sin^2\theta_{12} \gsim 0.335$. Nevertheless, both scenarios remain compatible with the AHEP global fit data. 

\begin{figure}[!h]
    \centering
    \includegraphics[width=0.48\linewidth]{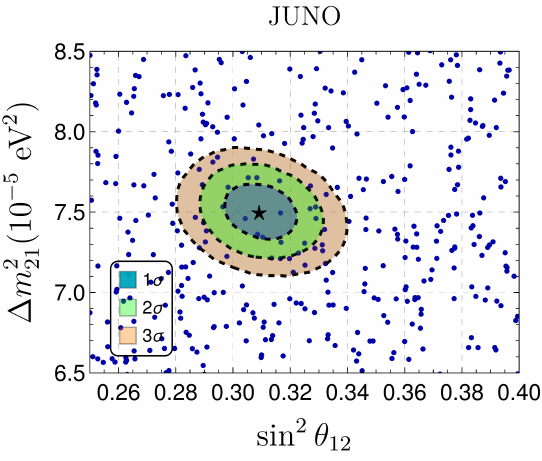}
    \includegraphics[width=0.48\linewidth]{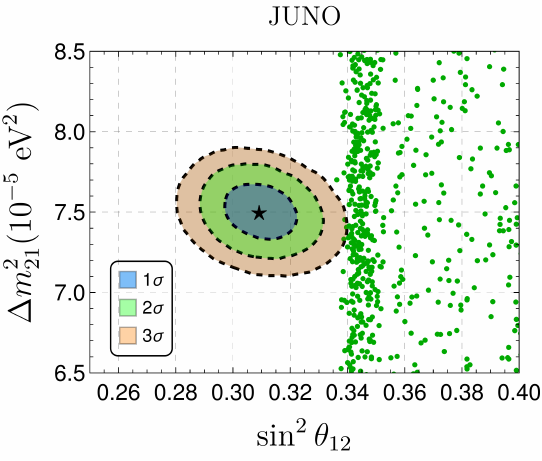}
    \caption{Correlation between the solar parameters $\sin^2\theta_{12}$ and $\Delta m^2_{21}$ predicted by the model, compared with the JUNO results~\cite{deSalas:2020pgw}. Case-I and case-II are shown in blue and green in the left and right panels, respectively.}
    \label{fig:th12m21_JUNO}
\end{figure}
However, the presence of this lower bound in case-II suggests a restricted compatibility with the experimental data. This observation becomes particularly compelling considering the recent JUNO measurement of $\sin^2\theta_{12}$. We therefore proceed to confront the allowed parameter space of the model with the JUNO data, which provides a more stringent and decisive test of the framework.
The allowed parameter space of the model and their comparison with the $1,2,3\sigma$ contours of JUNO are shown in Fig.~\ref{fig:th12m21_JUNO}. The corresponding best-fit value is indicated by a black star. As in Fig.~\ref{fig:th12m21_AHEP}, the left (right) panel corresponds to case-I (case-II).
Since the blue points span the entire allowed region in the $(\Delta m^2_{21}-\sin^2\theta_{12})$ plane, case-I remains a viable scenario. In contrast, for case-II, the model predicts a lower bound $\sin^2\theta_{12} \gsim 0.335$, causing all the green points to lie outside the $3\sigma$ JUNO contours. Consequently, in light of the JUNO results, this scenario of the model is nearly excluded.

Therefore, by comparing Figs.~\ref{fig:th12m21_AHEP} and \ref{fig:th12m21_JUNO}, we observe that prior to the JUNO measurements, case-II was also consistent and remained well within the $2\sigma$ range of the AHEP global fit data. However, once the JUNO results are taken into account, only the first scenario, case-I ($D=+\beta \frac{ u'}{\sqrt{3}}$), remains compatible, while the second one ($D=-\beta \frac{ u'}{\sqrt{3}}$) is almost ruled out.

\section{Conclusions}
\label{sec:conc}
We have developed a framework in which $\mu-\tau$ reflection symmetry is naturally realized as a consequence of an underlying $A_4$ flavor symmetry. Neutrino masses are generated via the type-II seesaw mechanism through the introduction of two $SU(2)_L$ triplet scalars: $\Delta_i$ and $\Delta'$, 
transforming as singlets and a triplet under $A_4$, respectively. The specific vev alignments of these scalars lead to an exact $\mu-\tau$ reflection symmetry in the neutrino sector. Depending on the choice of vev alignment for $\Delta'$, two distinct viable scenarios emerge.
An additional $Z_3$ symmetry ensures a diagonal charged lepton mass matrix, leaving the leptonic mixing entirely governed by the neutrino sector. The resulting $\mu-\tau$ reflection symmetry then fixes the atmospheric mixing angle to $\theta_{23}=45^\circ$ and the Dirac CP-violating phase to $\delta_{\rm CP}=\pm \pi/2$.

Beyond the well known predictions arising from $\mu-\tau$ reflection symmetry, our framework yields nontrivial and testable implications for the solar mixing angle $\theta_{12}$. In particular, when combined with the solar mass-squared difference $\Delta m^2_{21}$, the allowed parameter space of the model can be directly confronted with the latest JUNO measurements. Among the two distinct scenarios, one predicts $\sin^2\theta_{12} \gsim 0.335$, which is strongly disfavored by the JUNO results, although it remains compatible with the current AHEP global fit data. In both cases, $\sin^2\theta_{12}$ exhibits a robust correlation with the ratios of the underlying model parameters. 
 Imposing the neutrino oscillation constraints significantly narrows the allowed range of these ratios, thereby reducing the parametric freedom of the model and enhancing its phenomenological implications.

Furthermore, future measurements of the $\theta_{23}$ octant at DUNE and T2HK will play an important role in probing extensions of this framework that introduce controlled deviations from exact $\mu-\tau$ reflection symmetry. The observation of a non-maximal $\theta_{23}$ with a specific octant preference would provide compelling evidence in favor of such scenarios.

\section*{Acknowledgments}
%
\noindent
The author would like to acknowledge support from the National Research Foundation of Korea under grant NRF-2023R1A2C100609111. 

\appendix

\section{Electroweak Precision Observables: $S$, $T$, and $U$ Parameters} \label{app:stu_para}
In this section, we briefly discuss the corrections to the oblique parameters, namely $S$, $T$, and $U$ parameters~\cite{Peskin:1991sw}, induced by the presence of $SU(2)_L$ triplet scalars ($\Delta_i, \ \Delta'$). Within the type-II seesaw framework, these triplet scalars carry electroweak charges and therefore interact directly with the gauge bosons $W$ and $Z$. As a consequence, they contribute to the vacuum polarization amplitudes of $W$ and $Z$ bosons, leading to modifications in the oblique parameters.
The contributions of a scalar multiplet with general weak isospin and hypercharge to the oblique parameters have been computed in Ref. \cite{Lavoura:1993nq}. While their explicit evaluation in the type-II seesaw framework, involving a scalar triplet, can be found in Refs.~\cite{Chun:2012jw,Mandal:2022zmy}.
In our framework, the scalar sector is extended to include the multiple triplet scalars, each acquiring a vev as given in Eq.~\eqref{eq:vev}.
 The presence of such multiple triplet scalars modifies the corresponding expressions~\cite{Chun:2012jw,Mandal:2022zmy}, primarily through mixing among the triplet scalars. Nevertheless, their mixing with the Higgs doublet remains negligible due to the smallness of the triplet vevs, which are required for neutrino mass generation (see Eq.~\eqref{eq:range}).

These $SU(2)_L$ triplet scalars $\Delta_i$ and $\Delta'$ are assigned as singlets and triplet of $A_4$, respectively. In the $2 \times 2$ matrix notation, they can be expressed as 
\begin{align}
    \Delta_q \equiv \begin{pmatrix}
        \Delta^+_q/\sqrt{2} & \Delta^{++}_q \\
        \Delta^0_q & -\Delta^+_q/\sqrt{2}
    \end{pmatrix}, \ 
\end{align}
where $\Delta_q$ represents $\Delta_1,\Delta_2,\Delta_3,\Delta'_a,\Delta'_b,\Delta'_c$. After spontaneous symmetry breaking, these triplet scalars mix among themselves, neutral part with neutral and singly (doubly) charged with singly (doubly) charged scalars, respectively. We denote the resulting mass eigenstates as $H^0_i$, $H^+_j$, $H^{++}_k$, with corresponding masses $M^0_i$, $M^+_j$, $M^{++}_k$, respectively. The relation between gauge eigenstates and mass eigenstates is given by
\begin{align} \label{eq:mixtriplet}
    \Delta^0_q= U^0_{qi} H^0_i, \  \Delta^+_q= U^+_{qj} H^+_j, \  \Delta^{++}_q= U^{++}_{qk} H^{++}_k.
\end{align}
Taking into account the mixing relations defined in Eq.~\eqref{eq:mixtriplet}, and utilizing the general expressions for the oblique parameters of scalar multiplets derived in Ref.~\cite{Lavoura:1993nq}, we obtain the following expressions
\begin{widetext}
    %
\begin{align}
    &S=-\frac{1}{3 \pi} \sum_{p=-1}^1 \sum_{q,i} \left| U^{\boldsymbol{s_p}}_{qi} \right|^2 \left[ p \ln{M_{p,i}^2} + 6(p-s^2_W Q_p)^2 \xi \left(\frac{M_{p,i}^2}{M^2_Z}, \frac{M_{p,i}^2}{M^2_Z}\right) \right],
\\
       &T=\frac{1}{16 \pi s^2_W} \sum_{p=-1}^1 \sum_{q,i,j}  U^{\boldsymbol{s_p}}_{qi} U^{\boldsymbol{\ast s_{p-1}}}_{qj} \left[ (2+p-p^2) \eta \left( \frac{M_{p,i}^2}{M^2_W}, \frac{M_{p-1,j}^2}{M^2_W}\right) \right],
    \\
        &U= \frac{1}{6 \pi} \left[\sum_j\ln{(M_j^{+})^4}-\sum_i\ln{(M_i^{0})^2}-\sum_k\ln{(M_k^{++})^2} \right] \nonumber \\
        & \quad + \frac{1}{\pi} \sum_{p=-1}^1 \left[ \sum_{q,i,j}  U^{\boldsymbol{s_p}}_{qi} U^{\boldsymbol{\ast s_{p-1}}}_{qj}(-2-p+p^2) \xi \left( \frac{M_{p,i}^2}{M^2_W}, \frac{M_{p-1,j}^2}{M^2_W}\right)  + \sum_{q,i} \left| U^{\boldsymbol{s_p}}_{qi} \right|^2 2(p-s^2_W Q_p)^2 \xi \left( \frac{M_{p,i}^2}{M^2_Z}, \frac{M_{p,i}^2}{M^2_Z}\right) \right], 
    \end{align}
%
\begin{align}
 &\text{where} \quad   
    M_{p,i}=\{M^0_i,M^+_i,M^{++}_i \}, \ \boldsymbol{s_p}=\{0,+,++\} \ \text{and} \ Q_p=\{ 0,1,2 \} \ \text{for} \ p =\{ -1,0,1\}, \nonumber \\ &\text{and} \ s^2_W \equiv \sin^2{\theta_W}, \ \text{$\theta_W$ is the weak mixing angle.}  \nonumber 
    \end{align}
    \begin{align}
    & \xi(x,y)=\frac{4}{9}-\frac{5}{12}(x+y)+\frac{1}{6}(x-y)^2+\frac{1}{4}\left[x^2-y^2-\frac{1}{3}(x-y)^3-\frac{x^2+y^2}{x-y} \right]\ln{\frac{x}{y}}-\frac{1}{12}d(x,y)f(x,y), \nonumber \\
    & d(x,y)=-1+2(x+y)-(x-y)^2, \nonumber \\
    & 
f(x,y) =
\left\{
\begin{aligned}
& -2 \sqrt{d(x,y)} \left[
\arctan \left( \frac{x-y+1}{\sqrt{d(x,y)}} \right)
- \arctan \left( \frac{x-y-1}{\sqrt{d(x,y)}} \right)
\right], && \text{if } d(x,y) > 0 \\
& 0 &&  \text{if } d(x,y) = 0 \\
& \sqrt{-d(x,y)} \ln \left[
\frac{x+y-1+\sqrt{-d(x,y)}}{x+y-1-\sqrt{-d(x,y)}}
\right], && \text{if } d(x,y) < 0
\end{aligned}
\right. \\
& \eta(x,y)=x+y-\frac{2xy}{x-y}\ln{\frac{x}{y}} \ .
\end{align} \
\end{widetext}
In the above expressions, $M_{p-1,i}$ is formally undefined for $p = -1$, however, the corresponding terms vanish identically for $p = -1$ and therefore do not contribute.

In the limit of small mass splittings among the triplet components, the contributions to the $S,T,U$ parameters admit simplified expressions~\cite{Mandal:2022zmy}. While a detailed discussion of scenarios involving multiple triplet scalars is certainly of interest, it is beyond the scope of the current focus of the work. 
Nevertheless, it is important to note that such scalar sectors generically induce corrections to the $\rho$ parameter, thereby placing constraints on the vevs of the triplet scalars. For instance, in the minimal type-II seesaw scenario with a single triplet $\Delta$, one typically obtains the bound $v_{\Delta} \leq \mathcal{O}(1)$ GeV, which is expected to be even more restrictive in scenarios involving multiple triplets.
However, in our framework, the triplet vevs required for neutrino mass generation are naturally small. As evident from Eqs.~\eqref{eq:par} and \eqref{eq:range}, they remain sufficiently small even for $\mathcal{O}(1)$ Yukawa couplings. As a result, the induced corrections to the $\rho$ parameter and other electroweak observables are negligibly small, ensuring consistency with current experimental limits~\cite{ParticleDataGroup:2024cfk}.

\section{Tensor Product Rules of $\boldsymbol{A_4}$} \label{app:A4}

The $A_4$ symmetry is a non-abelian discrete flavor group. It corresponds to the group of even permutations of four objects and is isomorphic to the symmetry group of a regular tetrahedron. The group contains 12 elements and can be generated by two generators $S$ and $T$ obeying the relations:
\begin{equation} \label{eq:a4genrel}
S^2=T^3=(ST)^3=I.
\end{equation}
In the basis where $S$ and $T$ are real matrices, the generators are given by,
\begin{equation} \label{eq:a4genmat}
S=\left(
\begin{array}{ccc}
1&0&0\\
0&-1&0\\
0&0&-1\\
\end{array}
\right)\,, \quad
T=\left(
\begin{array}{ccc}
0&1&0\\
0&0&1\\
1&0&0\\
\end{array}
\right)\;.
\end{equation}
The $A_4$ group has four irreducible representations, three of them are one dimensional (i.e., three singlets) $1$, $1'$, and $1''$, and one of them is three dimensional (triplet) $3$.
The multiplication rules for these representations are given as follows:
\begin{eqnarray} \label{eq:a4mrule}
&1\times1&=1=1' \times 1'', \quad 1'\times 1'=1'', \quad 1''\times 1''=1',  \nonumber \\
&1 \times 3&=3, \quad 3\times 3= 1+ 1' + 1'' + 3_S + 3_A \quad .
\end{eqnarray}
where $3_S$ and $3_A$ denote the two independent triplet contractions arising from the product of two triplets.
If $a=\left(a_1,a_2,a_3\right)$ and $b=\left(b_1,b_2,b_3\right)$ are two triplets then their multiplication rules are constructed as follows~\cite{Ishimori:2010au}:
\begin{equation}\label{eq:pr}
\begin{array}{lll}
\left(ab\right)_1&=&a_1b_1+a_2b_2+a_3b_3\, ,\\
\left(ab\right)_{1'}&=&a_1b_1+\omega a_2b_2+\omega^2a_3b_3\, ,\\
\left(ab\right)_{1''}&=&a_1b_1+\omega^2 a_2b_2+\omega a_3b_3\, ,\\
\left(ab\right)_{3_S}&=&\left(a_2b_3+a_3b_2,a_3b_1+a_1b_3,a_1b_2+a_2b_1\right)^T\, ,\\
\left(ab\right)_{3_A}&=&\left(a_2b_3-a_3b_2,a_3b_1-a_1b_3,a_1b_2-a_2b_1\right)^T\,,
\end{array}
\end{equation}
where $\omega$ is the cube root of unity defined by $\omega = e^{2\pi i/3}$. The triplets $3_S$ and $3_A$ denotes the symmetric and antisymmetric contractions, respectively.

\section{Determination of Lepton Mixing Parameters} \label{app:lepmix}

The neutrino mass matrix given in Eq.~\eqref{eq:numatrix} can be diagonalized by a unitary transformation as follows
\begin{equation}
    U_\nu^T \mathcal{M}_\nu U_\nu = \text{diag}(m_1, m_2, m_3)\, ,
    \label{eq:diagnu}
\end{equation}
where $U_\nu$ is the unitary matrix and $m_{1,2,3}$ denote the physical neutrino mass eigenvalues. In the present framework, the charged lepton mass matrix is diagonal. Consequently, $U_\nu$ directly corresponds to the leptonic mixing Pontecorvo-Maki-Nakagawa-Sakata (PMNS) matrix,
\begin{equation}
    U_{\rm PMNS}\equiv U_{\nu}\, .
\end{equation}
We choose the symmetric parametrization of the lepton mixing matrix \cite{Schechter:1980gr,Rodejohann:2011vc}, given by

\begin{equation}
U_{\rm PMNS} = P(\delta_1, \delta_2, \delta_3) \, U_{23}(\theta_{23}, \phi_{23}) \, U_{13} (\theta_{13}, \phi_{13}) \, U_{12}(\theta_{12}, \phi_{12}) \, ,
\end{equation}
where $P(\delta_1, \delta_2, \delta_3)$ is a diagonal matrix of unphysical phases and the $U_{ij}$ are complex rotations in the $ij$ plane, as for example,
\begin{equation}
    U_{23} (\theta_{23}, \phi_{23}) = \left(\begin{matrix}
        1 & 0 & 0 \\
        0 & \cos\theta_{23} & \sin\theta_{23}\, e^{-i \phi_{23}} \\
        0 & -\sin\theta_{23} \,e^{i \phi_{23}} & \cos\theta_{23}
        \end{matrix} \right) \,.
\end{equation}
The phases $\phi_{12}$ and $\phi_{13}$ are relevant for neutrinoless double beta decay, while the combination $\delta_{CP} = \phi_{13} - \phi_{12} - \phi_{23}$ is the usual Dirac CP-violating phase measured in neutrino oscillations. The leptonic mixing angles can be extracted from the elements of the PMNS matrix through the following relations:
\begin{align}
    s_{13}= |U_{\rm PMNS}|_{13}, \quad  s_{12}= |U_{\rm PMNS}|_{12}/c_{13}, \quad  s_{23}= |U_{\rm PMNS}|_{23}/c_{13}.
\end{align}
where $s_{13,12,23}$ and $c_{13}$ denote the shorthand notations for $\sin{\theta_{13,12,23}}$ and $\cos{\theta_{13}}$, respectively.

\vspace{2.5cm}
\bibliographystyle{utphys}
\bibliography{references.bib} 
\end{document}